\documentclass{jfm}
\usepackage{graphicx}
\usepackage{epstopdf, epsfig}
\usepackage{tikz}

\usepackage{verbatim}
\usepackage{graphicx}
\usepackage{amsmath}%
\usepackage{longtable}%
\usepackage{bm}%
\usepackage{textcomp}
\usepackage{color}
\usepackage[normalem]{ulem}

\definecolor{mycol1}{rgb}{0.9047,0.1918,0.1988}
\definecolor{mycol2}{rgb}{0.2941,0.5447,0.7494}
\definecolor{mycol3}{rgb}{0.3718,0.7176,0.3612}
\definecolor{mycol4}{rgb}{1.0000,0.5482,0.1000}
\definecolor{mycol5}{rgb}{0.8650,0.8110,0.4330}
\definecolor{mycol6}{rgb}{0.6859,0.4035,0.2412}
\definecolor{mycol7}{rgb}{0.9718,0.5553,0.7741}
\definecolor{mycol8}{rgb}{0.6400,0.6400,0.6400}
\definecolor{mycol9}{rgb}{0.6365,0.3753,0.6753}

\newcommand{\colb}[1]{\textcolor{black}{#1}}

\shorttitle{Highly stratified compressible convection}
\shortauthor{J. P.  John and J. Schumacher}
\title{Compressible turbulent convection in highly stratified adiabatic background} 

\author{ John Panickacheril John\aff{1} 
\corresp{\email{ john.panickacheril-john@tu-ilmenau.de}}
and J\"org Schumacher\aff{1,2}
}
\affiliation{\aff{1} Institut f\"ur Thermo- und Fluiddynamik, Technische Universit\"at Ilmenau, Postfach 100565, D-98684 Ilmenau, Germany
\aff{2} Tandon School of Engineering, New York University, New York City, New York 11201, USA  }

\begin{document}
\maketitle

\begin{abstract}
Buoyancy-driven turbulent convection leads to a fully compressible flow with a prominent top-down asymmetry of first- and second-order statistics when the adiabatic equilibrium profiles of temperature, density and pressure change very strongly across the convection layer. The growth of this asymmetry and the formation of an increasingly thicker stabilized sublayer with a slightly negative mean convective heat flux $J_c(z)$ at the top of the convection zone is reported here by a series of highly resolved three-dimensional direct numerical simulations beyond the Oberbeck-Boussinesq and anelastic limits for dimensionless dissipation numbers, $0.1 \le D\le 0.8$, at fixed Rayleigh number $Ra=10^6$ and superadiabaticity $\epsilon=0.1$. The highly stratified compressible convection regime appears for $D > D_{\rm crit}\approx 0.65$, when density fluctuations collapse to those of pressure; it is characterized by an up to nearly 50\% reduced global turbulent heat transfer and a sparse network of focused thin and sheet-like thermal plumes falling through the top sublayer deep into the bulk.       
\end{abstract}


\section{Introduction}
Solar convection close to the surface proceeds in the presence of an extremely stratified adiabatic background equilibrium which is characterized by scale heights, the distances across which temperature $T$, density $\rho$, or pressure $p$ drop by an order of magnitude, as small as $\ell\sim 10^3$ km amounting to only 5\textperthousand$\,$ of the total depth of the solar convection layer, $H\sim 2\times 10^5$ km \citep{christ2002,nordlund2009solar,SKRMPhys2020}. It is one prominent example for a fully compressible convection flow which is highly asymmetric when up- and down-welling thermal plumes are compared. Driven by the strong radiative cooling at the top surface, cold plasma sinks into the interior ($u_z<0$) with nearly the speed of sound at the narrow edges of the granules, the convection cells in the solar case. Plasma rises moderately to the surface across the granule interior with a typical diameter of about $10^3$ km \citep{magic2013a}. It is yet open, how deep the narrow plumes reach into the highly stratified interior \citep{CossetteRast2016,BrandenburgApJ2017,AndersApJ2019} and how much they get focused by compressibility effects to overcome turbulent dispersion. Answers to these questions would also provide insights on the vertical extent of supergranules, the larger-scale convection cells in the background with an extension of about 30 granule diameters \citep{Rincon2018,HansonAA2020} and on the related {\em solar conundrum} of anomalously low velocity fluctuation amplitudes from helioseismology in comparison to models \citep{HanasogePNAS2012}. All this sets the physical motivation for the present study.

In this work, we want to study fully compressible convection in the presence of highly stratified adiabatic equilibrium profiles for $T$, $\rho$, and $p$. We thus analyse a series of direct numerical simulations (DNS) towards the high stratification limit of $D\to 1$ for the dissipation number $D$ (defined below). Our goal is to analyse the genuine compressibility effects isolated from complex multi-physics, as in the solar example above, and from temperature dependencies of dynamic viscosity and thermal conductivity. Our three-dimensional DNS beyond the Oberbeck-Boussinesq \citep{CJ2012} and anelastic regimes \citep{OP1962,VerWSAJ2015,jones2022} lead to strongly differing boundary layer dynamics at the top and bottom, such as a sublayer formation at the top with a slightly negative mean convective heat flux, $J_c(z)\sim \langle u_z T^{\prime}\rangle_{A,t}$, which causes a reduction of the global turbulent heat transfer by almost 50\% for $D=0.8$. In this layer, narrow focused thermal plumes form at the top and fall deeply into the highly stratified bulk of the layer. Our study thus demonstrates that some typical phenomena in natural convection flows can be understood in significantly simpler compressible flow configurations even though being far away from the natural flow case in terms of Rayleigh and Prandtl numbers.

\section{Simulation model and parameters of compressible convection}

The equations of motion for compressible convection are given by   
\begin{align}
\partial_t \rho  + \partial_i (\rho u_{i}) &= 0
\label{eq:mass}\,,\\
\partial_t (\rho u_{i}) + \partial_j (\rho u_{i} u_{j})  &= -\partial_i p  + \partial_j \sigma_{ij} + \rho g \delta_{i,3}
\label{eq:mom}\,,\\
\partial_t (\rho e) + \partial_j (\rho e u_{j} ) &= -p \partial_i u_{i} +
\partial_i(k \partial_i T) + \sigma_{ij} S_{ij}
\label{eq:ener}\,,\\
p &= \rho R T \quad\textrm{ where }\quad R= C_{p} -C_{v} .\end{align}
These equations correspond to mass, momentum and energy conservation laws along with the ideal gas equation of state. Here, $\rho$, $\rho u_{i}$, $p$, $\rho e$, $T$ are the mass density, momentum density components, pressure, internal energy density, and temperature, respectively. The viscous stress tensor is ${\bm \sigma}=2\mu{\bm S}+ 2\mu{\bm I}({\bm \nabla}\cdot {\bm u})/3$ with the Kronecker tensor ${\bm I}$ and the rate of strain tensor ${\bm S} = ({\bm \nabla} {\bm u}+{\bm \nabla} {\bm u}^T)/2$. The dynamic viscosity $\mu$ is assumed to be constant in these simulations. The thermal conductivity $k$ is related to the viscosity through the Prandtl number, $k= \mu C_{p}/Pr$. In our DNS, $Pr= 0.7$. $C_{p}$ and $C_{v}$ correspond to specific heat at constant pressure and volume, respectively. Their ratio, $\gamma= C_{p}/C_{v}= 1.4$ for a diatomic gas. \colb{The internal energy is defined as $e=C_{v}T$}.

A uniform grid is used in $x$-- and $y$--directions along with periodic boundary conditions. In wall-normal $z$--direction, a non-uniform grid with a point clustering near the walls is taken, which follows a hyperbolic tangent stretching function. Spatial derivatives are calculated by a 6th-order compact  scheme  for all points except near the walls \citep{lele1992,BDBjfm2022}; there 4th- and 3rd--order compact schemes are used at the last two grid points near the wall. No-slip, isothermal boundary conditions are applied at the top and bottom.  The boundary condition for $p$ is evaluated using the $z$-component of the momentum equation at $z=0, H$, $\partial p/\partial z =  \partial \sigma_{iz}/\partial x_{i} + \rho g$. The fields are advanced in time by a low storage 3rd-order Runge-Kutta with a Courant number of ${\rm CFL}=0.5$.

\begin{table}
\begin{center}
\begin{tabular}{ccccccccccc}
Case & $D$ & $T^{B}_{a}/T^{T}_{a}$ & $\rho^{B}_{a}/\rho^{T}_{a}$  & $M_{t}^{\rm max}$ & $Nu$ & $Re$  & $\lambda_{\rho_{\rm min}}$  & $\delta_{u_{z}T}$ & $ \left( \lambda^{\rm max}_{\rho' T'},\delta_{\rho' T'}     \right)$  &   Style  \\ 
\hline
    1 & $0.10$ & $1.10$ & $1.3$ &  $0.14$ & $7.94 \pm 0.05$     & $424 \pm 2 $ & 0.002 & 0.001 & $\left(0.001, 0 \right)$ & 
    \begin{tikzpicture}
\draw[mycol1] (-0.5,0) -- (0.5,0);
        \end{tikzpicture}    \\
    2 & $0.34$ & $1.52$ & $2.8$ & $0.32$   &  $7.64 \pm 0.01 $  & $414 \pm 1 $ & 0.012 & 0.005 & $\left( 0.006, 0.012 \right)$ &
    \begin{tikzpicture}
\draw[mycol2] (-0.5,0) -- (0.5,0);
        \end{tikzpicture}     \\
3 & $0.50$ & $2.00$ & $5.7$ & $0.41$  & $6.93 \pm 0.02$   & $407 \pm 2$ & 0.024 & 0.010 & $\left( 0.012, 0.024 \right)$ &
\begin{tikzpicture}
\draw[mycol3] (-0.5,0) -- (0.5,0);
        \end{tikzpicture}     \\
4 & $0.60$ & $2.50$ & $9.9$ & $0.49$  & $6.24 \pm 0.01$   & $370 \pm 8 $ & 0.037 & 0.016 & $\left( 0.018, 0.037 \right)$ &
\begin{tikzpicture}
\draw[mycol4] (-0.5,0) -- (0.5,0);
        \end{tikzpicture}     \\
5 & $0.65$ & $2.86$ & $13.8$ & $0.55$  & $5.90 \pm 0.01$    & $364 \pm 2$ & 0.042 & 0.019 & $\left( 0.021, 0.042 \right)$  &
\begin{tikzpicture}
\draw[mycol5] (-0.5,0) -- (0.5,0);
        \end{tikzpicture}     \\
 6 & $0.70$ & $3.33$ & $20.3$ & $0.54$  & $5.80 \pm 0.04$  & $425 \pm 2$& 0.051 & 0.023 & $\left( 0.023, 0.051 \right)$ & 
\begin{tikzpicture}
\draw[mycol6] (-0.5,0) -- (0.5,0);
        \end{tikzpicture}     \\ 
7 & $0.75$ & $4.00$ & $32.0$ & $0.60$  &  $5.14 \pm 0.01$  & $399 \pm 0 $ & 0.068 & 0.033 & $\left(0.028, 0.068 \right)$ &
\begin{tikzpicture}
\draw[mycol7] (-0.5,0) -- (0.5,0);
        \end{tikzpicture}     \\
8 & $0.80$ & $5.00$ & $55.9$ &  $0.61$ & $4.46 \pm 0.03$   & $356 \pm 1 $ & 0.098 & 0.051 & $\left(0.035, 0.098 \right)$ &
\begin{tikzpicture}
\draw[mycol8] (-0.5,0) -- (0.5,0);
        \end{tikzpicture}     \\
\end{tabular}
\end{center}
\caption{
\label{tab:simu1} 
List of the direct numerical simulations. All cases have $Pr= 0.7$, $Ra \approx 10^6$, $\epsilon= 0.1$, $\gamma= C_{p}/C_{v}= 1.4$, and an aspect ratio, $\Gamma=L/H=4$ resolved by $N_x:N_y:N_z=512: 512: 256$ grid points. We list dissipation number $D$, bottom-to-top ratios of adiabatic temperatures and densities, the global maximum  turbulent Mach number $M_t^{\rm max}$, the Nusselt and Reynolds numbers, $Nu$ and $Re$. The error bars of $Nu$ and $Re$ are less than $2 \%$ for  all cases. Furthermore, we display boundary layer scales for density, convective heat flux, and density-temperature correlations. In all figures that follow, coloured lines are given as in this table.}
\end{table}
%%%%%%%%%%%%%%%%%%%%%
\begin{figure}
\hspace{-15mm}
\includegraphics[clip,trim={0mm 0mm 0mm 0mm},width=1.2\textwidth]{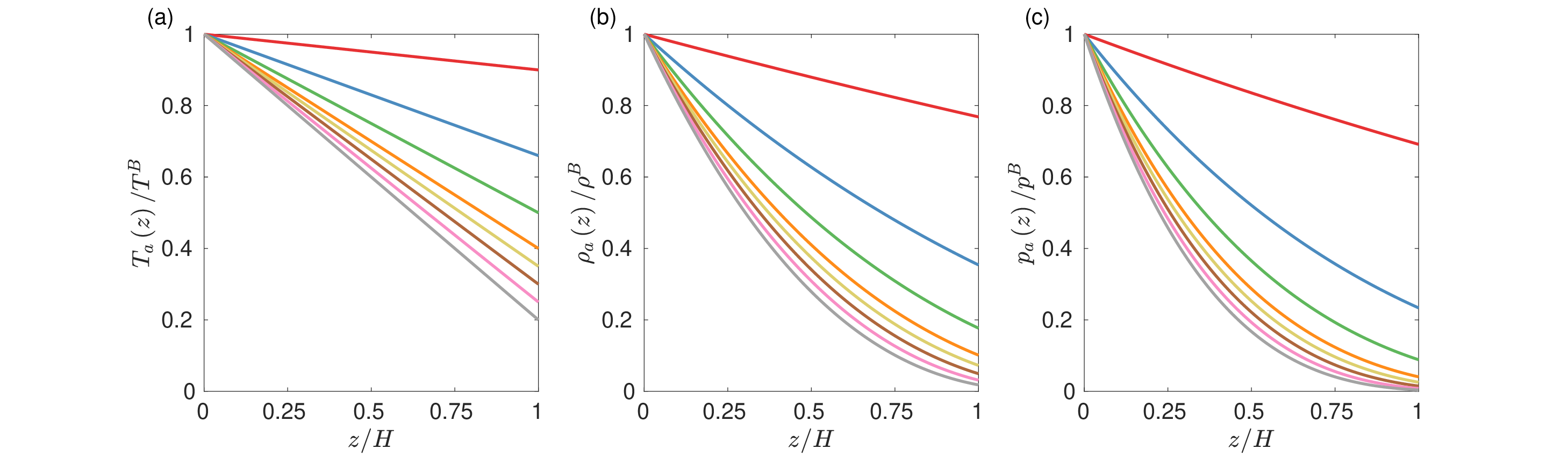}
\caption{\label{fig:Fig0} Adiabatic equilibrium profiles of the thermodynamic state variables. (a) Temperature $T$, (b) density $\rho$, and (c) pressure $p$. The difference between top and bottom values for all three variables increases with $D$, see table \ref{tab:simu1} for the color code description.}  
\begin{picture}(0,0)
\end{picture}
\end{figure}

Incompressible Rayleigh-B\'{e}nard convection is determined by the Rayleigh and Prandtl numbers, $Ra$ and $Pr$. In compressible convection, additional dimensionless parameters need to be introduced. The first one is the {\em dissipation number} $D$ \colb{\citep{VerWSAJ2015,jones2022} } which is given by 
\begin{equation}
    D= \frac{gH}{C_{p}T^{B}}\,. 
\end{equation}     
The parameter $D$ can be considered as a rescaled dry adiabatic lapse rate as it determines differences in the thermodynamic properties across the layer depth in a purely isentropic process. The adiabatic temperature ratio between top (T) and bottom (B) is  $T^{T}_{a} /T^{B}_{a}=( 1 - D)$, the corresponding density ratio follows to $\rho^{T}_{a} /\rho^{B}_{a} = \left[ T^{T}_{a} /T^{B}_{a} \right]^{1/ \left( \gamma -1 \right) }$. The subscript "a" corresponds to the adiabatic equilibrium without convection. \colb{The equilibrium profiles for all state variables and all values of $D$ are shown in figure \ref{fig:Fig0}}. In order to initiate convective motion, the actual temperature gradient across the layer must be greater than the one due to a purely adiabatic process \citep{jeff1930}, i.e., $T^{B} - T^{T} > T_{a}^{B} - T_{a}^{T}$. In our study, the bottom plate is taken as the reference, thus $T^{B}= T_{a}^{B}$ is a constant for all cases; the instability criterion reduces to $T^{T} < T_{a}^{T}$.   

The second additional compressible convection parameter is the {\em superadiabaticity} $\epsilon$ \colb{\citep{VerWSAJ2015,jones2022} }, which is given by 
\begin{equation}
    \epsilon= \frac{T_{a}^{T} - T^{T}}{T^{B}}\,.
\end{equation}
Superadiabaticity is the excess relative temperature gradient with respect to the adiabatic equilibrium  state, or in other words, a  measure of the driving of convection. If one takes the characteristic free-fall velocity, $U_{f}= \sqrt{\epsilon g H} $, then a free-fall Mach number,
\begin{equation}
M_{f}(z) = \frac{U_{f}}{\gamma R \langle T(z) \rangle_{A,t}} = \sqrt{\frac{\epsilon D}{\gamma - 1}} \sqrt{\frac{T^{B}}{\langle T(z)\rangle_{A,t}}} \,,
    \label{eq:mach}
\end{equation}
follows. \colb{The notation $\langle \cdot\rangle_{A,t} $  corresponds to a combined average over the horizontal directions and time which is given for a field $X$ as 
\begin{equation}
    \langle X (z) \rangle_{A,t} = \frac{ \sum^{N_{t}}_{m=1} 
    \sum^{N_{x}}_{i=1} \sum^{N_{y}}_{j=1} X(x_i, y_j,z,t_m)}{N_{x}N_{y}N_{t}},
\end{equation}
where $N_{x}$, $N_{y}$, and $N_{t}$ are the grid point numbers with respect to $x$-, $y$-directions, and number of snapshots, respectively.}
The strength of compressibility is thus a function of both, $\epsilon$ and $D$. From \eqref{eq:mach}, it is expected that compressibility at the top is higher than at the bottom. For $\epsilon \rightarrow 0$,  we get the anelastic approximation where the acoustic waves are filtered out from the equations. Rayleigh-B\'ernard convection in the Oberbeck-Boussinesq (OB) regime follows for $\epsilon, D \rightarrow 0$. On the basis of the new parameters, we will use $Ra= \epsilon g H^{3}/(\nu^{B} \kappa^{B})$ and Prandtl number, $Pr= \nu^B/\kappa^B$, where $\kappa^B= k/\left(C_{p} \rho_a^B \right)$ is the temperature diffusivity and $\nu^B=\mu/\rho_a^B$ the kinematic viscosity. \colb{The Rayleigh number is here defined with respect to the bottom variables and thus constant for all cases. Unlike $Ra$, which varies across depth due to density dependence, Prandtl number $Pr$ is  a constant in the entire domain, since the density drops out for $\nu$ and $\kappa$.} In this work, we focus on a constant and low but finite superadiabaticity, $\epsilon=0.1$ for a range of $D$ from $0.1$ to $0.8$. All other governing parameters such as the, $Ra \approx 10^6$, $Pr= 0.72$ and aspect ratio, $\Gamma= 4$ are kept constant, see  table \ref{tab:simu1}. In all figures that follow, unless specified otherwise, we use the plotting style given in the last column. \colb{From table \ref{tab:simu1}, we observe that in our study, the strength  of density stratification $\left(\rho_{a}^{B}/\rho_{a}^{T}\right)$ ranges from 1.3 at $D= 0.1$ to 55.9 at $D=0.8$. The degree of stratification in previous compressible convection simulations is thus significantly smaller than what we will study here. The strongest stratification ratios in \cite{VerWSAJ2015}, \cite{jones2022},  and \cite{tilgner2011} are 20.1, 10 and  4.6, respectively.}

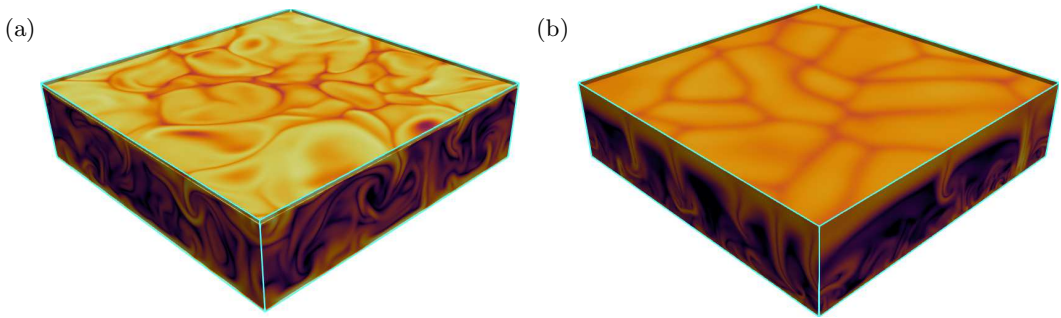
\begin{figure}
\begin{center}
\includegraphics[clip,trim={140mm 20mm 195mm 20mm},width=0.47\textwidth]{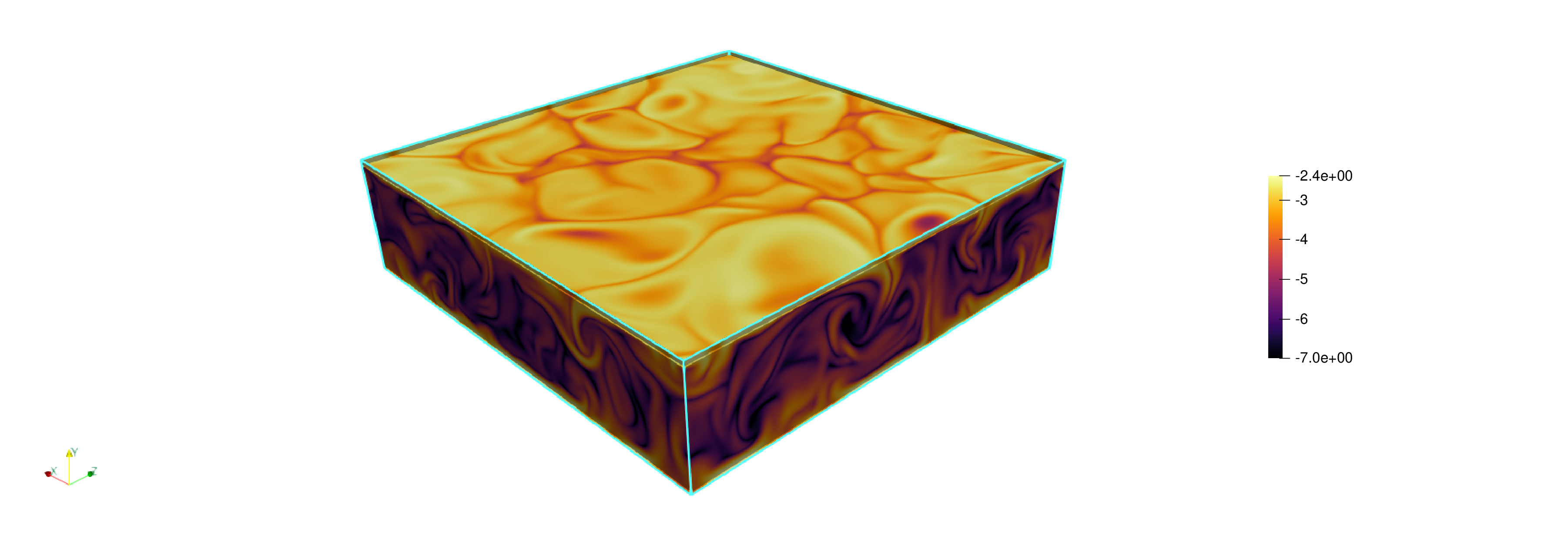}
\includegraphics[clip,trim={140mm 12.5mm 167.5mm 25mm},width=0.52\textwidth]{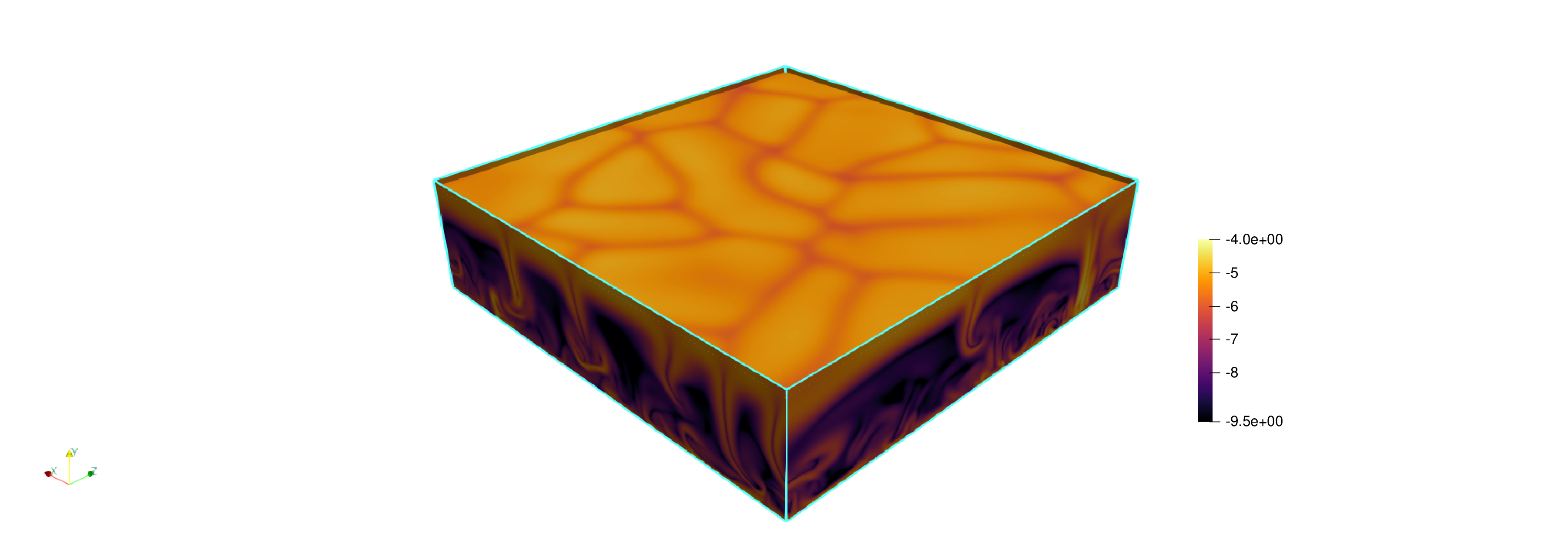}
\caption{\label{fig:Fig1} Visualization of the plume structure of the superadiabatic temperature field $T_{sa}$. Contours of $\ln |{\bm \nabla} T_{sa} |$ are shown at two instants, (a) for case 1 with $D=0.1$ and (b) for case 8 with $D=0.8$. Minimum/maximum contour levels  correspond to -7.0/-2.4 in (a) and -9.5/-4.0 in (b). The top contour surface is slightly below $z=H$.}  
\begin{picture}(0,0)
\put(-205,160){(a)}
\put(-5,160){(b)}

\end{picture}
\end{center}
\end{figure}

\section{Growing asymmetry with increased stratification}
The adiabatic temperature profile, $T_{a}(z)=T^B_a(1-Dz)$, is the equilibrium profile. Thus the part of temperature corresponding to convection is the superadiabatic temperature defined as  
\begin{equation}
    T_{sa} ({\bm x},t)= T({\bm x},t) - T_{a}(z)\,.
\end{equation}
In figure \ref{fig:Fig1}, we compare the contour visualization of  $ \ln \left(| \mathbf{\nabla} T_{sa}| \right)$ for the two extreme cases in our series with $D= 0.1$ (low)  and $D= 0.8$ (high). Panel (a) corresponds to case 1, closest to the OB limit. Not surprisingly, the plumes from the top and bottom are more or less symmetric in shape and frequency.  However, this top-bottom-symmetry disappears completely for the highly stratified case 8 in panel (b). One observes a network of slender thin plumes \citep{rast1998} originating from the top boundary layer that fall deep into the bulk. Recently, \cite{JPSc2023} showed that thinner plumes from the top boundary are a common characteristic of compressible convection regardless of the relative boundary layer thickness, but here the asymmetry is very pronounced.

\begin{figure}
\includegraphics[clip,trim={0mm 0mm 0mm 0mm},width=0.5\textwidth]{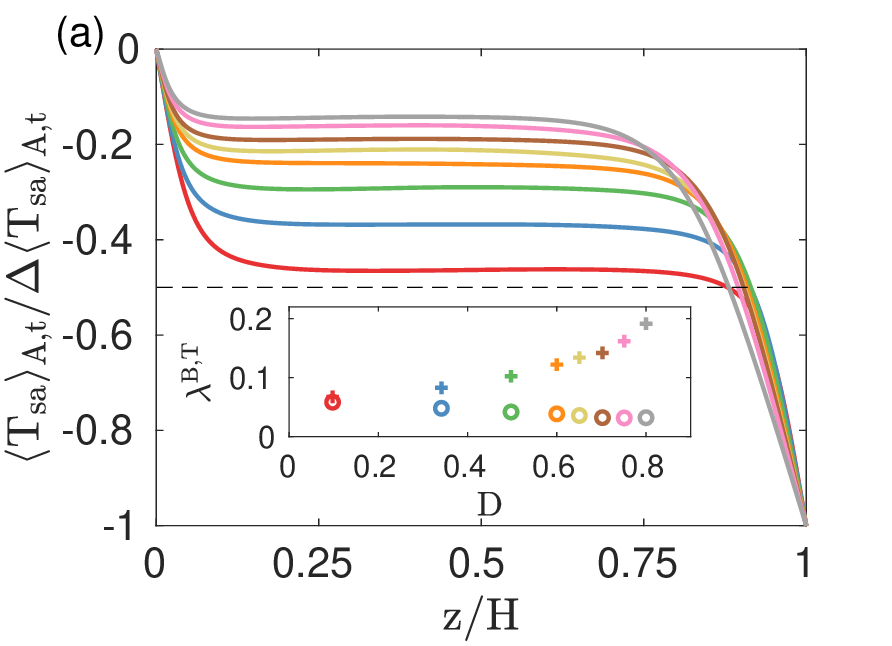} \includegraphics[clip,trim={0mm 0mm 0mm 0mm},width=0.5\textwidth]{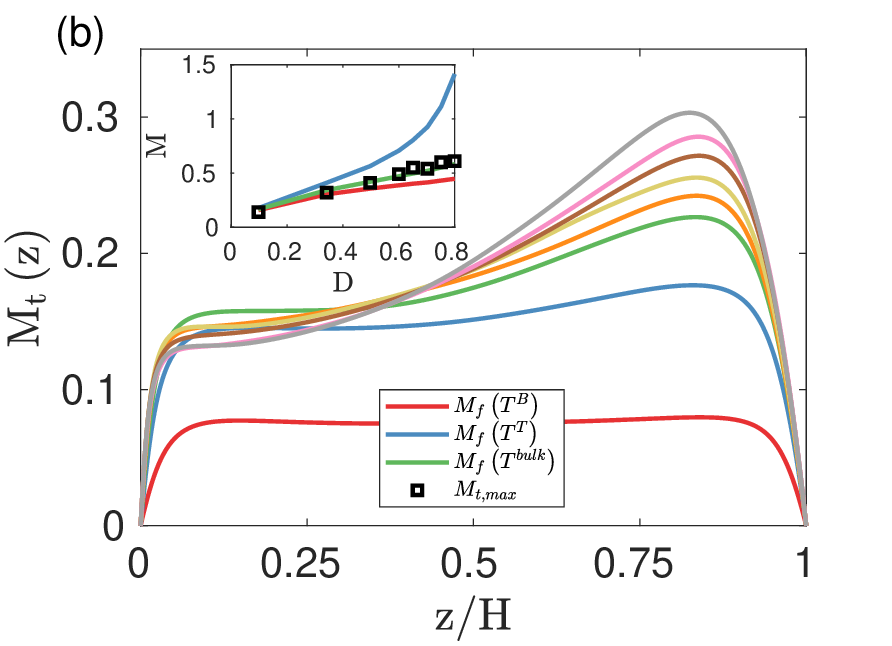} 
\caption{\label{fig:Fig2} Asymmetry of the (a) normalized  superadiabatic temperature and (b) the turbulent Mach number $M_t$. The inset in (a) corresponds to mean thermal boundary layer thickness $\lambda^{B,T}(D)$. Symbols $+$ and $\rm o$ are for the top and bottom boundary layer thickness, respectively. The inset in (b) shows $M_f(D)$ from \eqref{eq:mach}. The red, blue, and green lines are for $z/H = 0, 1$, and 0.5, respectively. The black squares are the maximum Mach numbers as given table \ref{tab:simu1}, which are observed in the simulations over the whole space and time. The legend in (b) corresponds to the inset figure.} 
\begin{picture}(0,0)
\end{picture}
\end{figure}

Similar asymmetry is observed for the mean profile of the superadiabatic temperature, $\langle T_{sa} (z) \rangle_{A,t}$. In figure \ref{fig:Fig2}(a), we plot the normalized superadiabatic temperature, $\langle T_{sa}(z)\rangle_{A,t}/ \Delta \langle T_{sa} \rangle_{A,t}$  as a function of depth $z$ where $\Delta \langle T_{sa} \rangle_{A,t}=\langle T_{sa}(z=0) \rangle_{A,t} - \langle T_{sa}(z=H) \rangle_{A,t}$.  Clearly one observes that the asymmetry, i.e., the offset from 0.5 and thickness of the top boundary layer in comparison to the bottom one, increases considerably with growing dissipation number. More detailed, as $D$ increases, the  bulk temperature is increasingly closer to the  bottom temperature, \colb{consistent with  \citet{VerWSAJ2015,jones2022,tilgner2011}. Differently, the experiments of \cite{WL1991} showed that the bulk temperature is closer to the prescribed top plate value with $\left(\nu^{T}, \kappa^{T} \right) < \left( \nu^{B}, \kappa^{B} \right)$. This results in a thinner top boundary layer and a bulk temperature closer to that at the top. Due to strong density stratification, we get $\left(\nu^{T}, \kappa^{T} \right) > \left( \nu^{B}, \kappa^{B} \right) $ and thus thicker layers at the top, which is detailed later. A trend similar to \cite{WL1991} is observed for $\epsilon / D \gg 1 $ in  \cite{JPSc2023}.} 

%%%%%%%%FigureAlert%%%%%
%%%%%%%%%%%%%%%%
We can quantify this thickness as the mean of a {\em local} thermal boundary layer thickness \citep{JSJSJFM2014} which is given by
\begin{equation}
    \tilde{\lambda}^{B,T}(x,y)= \frac{1}{H}\left |
\langle T_{sa} \left( 0,H \right) - T_{sa} \left( H/2 \right)    \rangle_{A,t} \right|   
\left |      
\frac{\partial \langle T_{sa} \rangle_{A,t}  }{\partial z} \right|^{-1}
_{z=0,H}. 
\end{equation}
We plot $\lambda^{B,T}=\langle \tilde{\lambda}^{B,T}\rangle_{A,t}$ as a function of $D$ in the inset of figure \ref{fig:Fig2}(a). The difference between the boundary layer thickness at the bottom and top plate increases with $D$ in agreement with the differences observed in the distributions of the instantaneous, local thermal boundary layer thicknesses \citep{JPSc2023}. The top boundary layer thickness increases drastically with $D$ whereas bottom boundary layer thickness is almost invariant.  This along with increased asymmetry in the bulk towards the bottom plate, as seen in the normalized $T_{sa}$-profiles, suggests that at high $D$, the mixing is not as efficient as in OB and compressibility effects are mainly localized near the top boundary.

An asymmetry similar to that of $\langle T_{sa}(z)\rangle_{A,t} $ is observed for the turbulent Mach number too with significant localized compressibility seen near the top boundary. The turbulent Mach number is defined as  
\begin{equation}
    M_{t}(z) = \frac{u_{\rm rms}(z)}{\sqrt{\gamma R \langle T(z) \rangle_{A,t}}}\quad \mbox{with} \quad  u_{\rm rms}(z)= \sqrt{ \sum_{i=1}^3 \left [\langle u_{i}^{2}(z)\rangle_{A,t} - \langle u_{i}(z) \rangle^{2}_{A,t}\right ]}\,.
\end{equation}
The turbulent Mach number at the top increases monotonically with $D$, whereas the behaviour at the bottom boundary is non-monotonic with $D$. At the bottom, $M_{t}$ increases up to $D= 0.5$ and then decreases with $D$ as a result of the increased asymmetry between the boundary layers, see figure \ref{fig:Fig2}(b). 

In the inset of figure \ref{fig:Fig2}(b), we compare the resulting maximum  turbulent Mach number in the flow with the  estimate from \eqref{eq:mach}. The black squares are the maximum local  turbulent Mach number for each $D$ encompassing  all time steps and spatial locations. The  red and blue lines correspond to $M_{f}$-estimates using the bottom and top temperatures, $T^B$ and $T^T$, respectively. One finds that the resulting maxima of $M_t$ are in between these estimates. This is due to no-slip boundary conditions at the wall; thus the velocity fluctuations peak only at the outer region of the boundary layer and in the bulk. Since the (mean) bulk temperature is closer to that at the bottom plate, the values of $M_t$ are closer to $M_{f} \left(T^{B} \right)$. A really close agreement for the simulations is seen,  when the Mach number is estimated using the bulk temperature, green line in the inset figure. This suggests that a simple estimate such as in \eqref{eq:mach} can give us a good idea about the level of compressibility in the flow. 

We would like to point out that in compressible turbulence $M_{t}$ is not sufficient to characterize the system; an additional parameter $\delta$, the ratio of dilatational to solenoidal root mean square velocities, needs to be included as discussed in \cite{DJ2020,JDS2019,JDScst2020,JDSJFM2021}. \colb{This is not surprising as in compressible convection two extremely different conditions,  $\epsilon/D \ll 1$ and $\epsilon/D \gg 1$, can lead to the  same $M_t$ despite different dynamics. \cite{JPSc2023} showed that an increasing $\epsilon$ leads to enhanced heat transfer, whereas it is decreased for increasing $D$ at low $\epsilon$. Even though $M_t$ would be comparable for both cases, $\delta$ will differ in these limits. A similar conclusion, that the Mach number along with a measure of dilatation in the flow field is required for the statistical analysis, was reported for compressible turbulent channel flows \citep{BDBaJFM2023,BDBaiaa2023}.}

\section{Fluctuations of state variables and regime transition}
\begin{figure}
\includegraphics[clip,trim={0mm 0mm 0mm 0mm},width=0.46\textwidth]{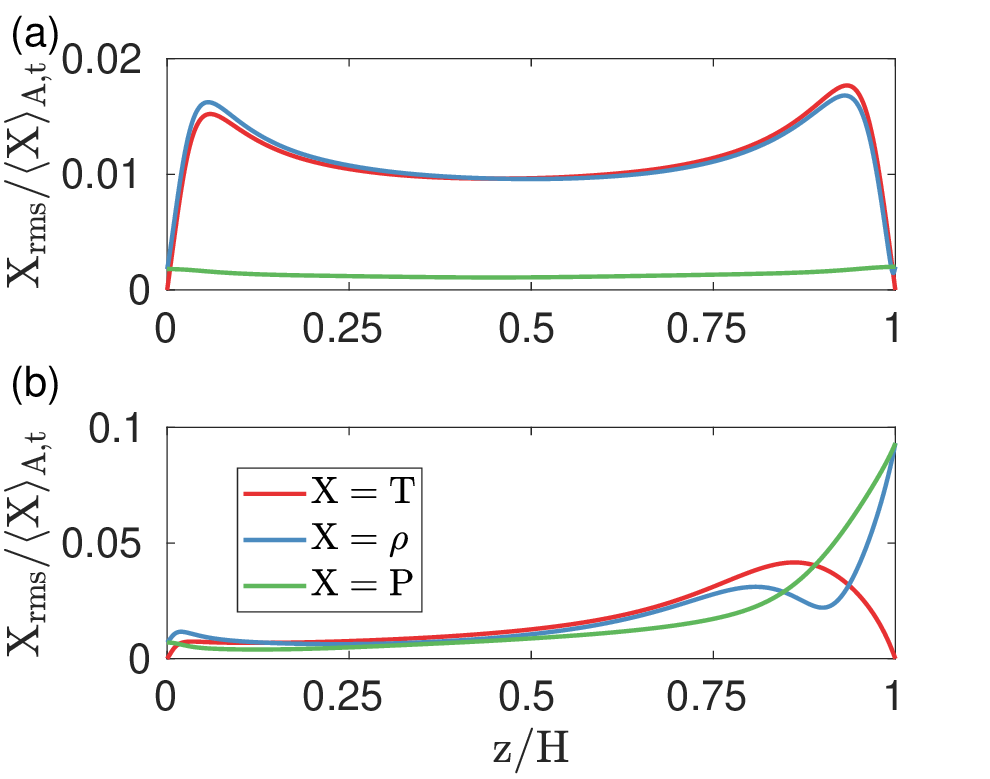} \includegraphics[clip,trim={0mm 0mm 0mm 0mm},width=0.46\textwidth]{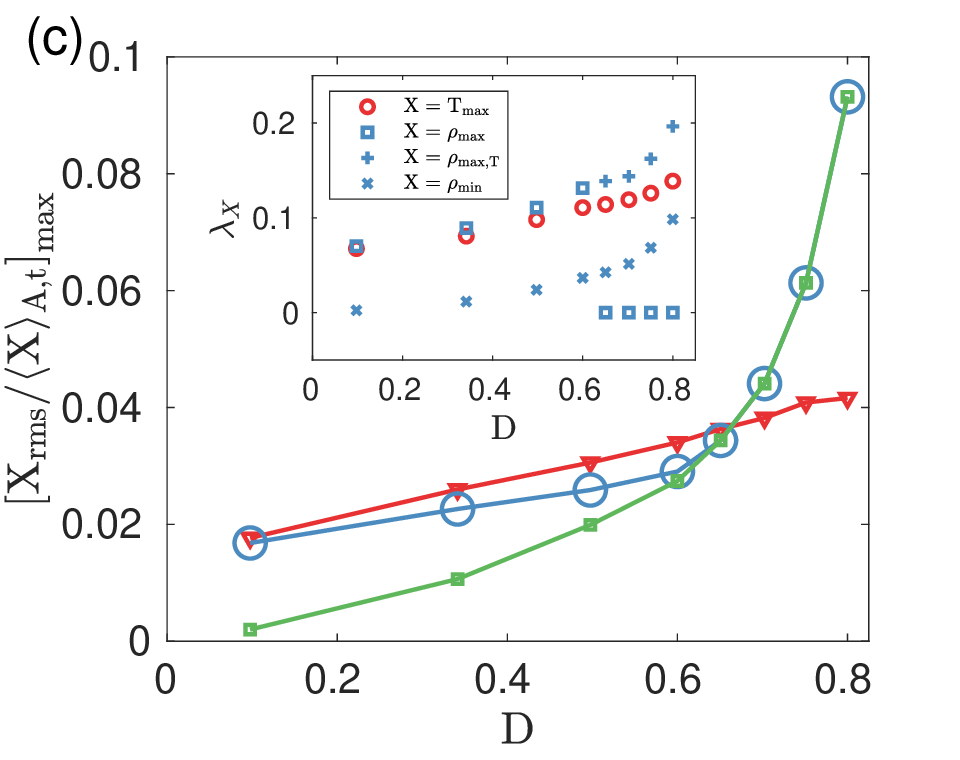} 
\caption{\label{fig:Fig3} Relative fluctuations of  thermodynamic quantities for (a) $D = 0.1$ and (b) $D= 0.8$. (c) Maximum of the  relative thermodynamic fluctuations versus dissipation number, $D$. For all figures, the red, blue and green correspond to temperature, density and pressure respectively. Inset of (c) shows four different distances $\lambda_{X}$ as a function of $D$. For the definition of the specific $\lambda_{X}$ we refer to the text in section 4 and the legend of the inset.} 
\begin{picture}(0,0)
\end{picture}
\end{figure}

A major consequence of compressibility are the fluctuations of $p$, $\rho$, and $T$. In figure \ref{fig:Fig3}, we compare these fluctuations as a function of depth for the two extreme cases of our DNS series, $D= 0.1 $ and $0.8$. \cite{JPSc2023} showed that $D\le 1-\epsilon$, which would result in a maximally possible theoretical value of $D=0.9$ in our DNS series. The relative fluctuations are defined as $X_{\rm rms}(z)/\langle X(z) \rangle_{A,t}$ with $X_{\rm rms} = (\langle X^2 \rangle_{A,t} - \langle X \rangle_{A,t}^2)^{1/2}$. Panel (a) at $D=0.1$ shows that pressure fluctuations are negligible. The magnitude of the density and temperature fluctuations are of the same order of magnitude and the profiles of both quantities are practically identical. These findings are consistent with the OB limit at $\epsilon, D \rightarrow 0$.

In figure \ref{fig:Fig3}(b), the relative fluctuations of $p$, $\rho$, and $T$ at $D= 0.8$ are shown. As a result of high compressibility at $D=0.8$ at the top, see figure \ref{fig:Fig2}(b) for $M_t(z)$, the thermodynamic fluctuations are considerably higher compared to the bottom one. Close to the top boundary, the pressure and density fluctuations dominate the temperature fluctuations. This is due to the constant temperature imposed  at the boundaries which results in a isothermal condition. For an assumed ideal gas, $p\propto \rho$. Indeed this is evident for $D= 0.8$ at the top boundary as the relative fluctuations of density and pressure perfectly match at the top boundary. This would be also true at the bottom boundary and even for $D=0.1$, but it is not evident due to low $M_{t}$. Compressibility effects imply in the following discussion that we face a large magnitude of pressure fluctuations.

As we  move away from the wall, fluctuations of both, $p$ and $\rho$, decrease, but are not perfectly correlated anymore due to the growing temperature fluctuations. After a  critical distance from the wall, pressure fluctuations continue to decrease; the density fluctuations reach a minimum at $z=z_{\ast}$. This local minimum of $\rho_{\rm rms}(z)/\langle\rho(z)\rangle_{A,t}$ exists when pressure and temperature fluctuations become comparable in magnitude. For the highest dissipation number $D= 0.8$, density fluctuations at $z<z_{\ast}$ are driven by temperature fluctuations in contrast to $z>z_{\ast}$.  One also notes that the location of the secondary maximum of $\rho_{\rm rms}(z)/\langle\rho(z)\rangle_{A,t}$ does not match the local maxima of $T_{\rm rms}(z)/\langle T(z)\rangle_{A,t}$. This implies that pressure fluctuations, although not dominant,  remain significant and thus compressibility effects too \citep{jones2022}. 

%%%%%%%%Figure3Alert%%%%%
%%%%%%%%%%%%%%%%

These features are absent for $D= 0.1$ in panel (a). Thus there seems to exist a critical $D$ where a transition to the compressibility-dominated convection regime exists that is qualitatively different to the OB or anelastic limits. In order to identify the critical $D$, we plot the maximum relative thermodynamic fluctuations with respect to $D$ in figure \ref{fig:Fig3}(c). We find a $D_{\rm crit}\approx 0.65$ for the present $(Ra,\epsilon)$.  For $D < D_{\rm crit}$, maxima of density and temperature fluctuations dominate over those of the pressure fluctuations. As $D$ increases, the maximum pressure fluctuations start to increase as well which is connected with increasing differences in the maxima of density and temperature fluctuations. For $D \ge D_{\rm crit}$, the maxima of pressure and density fluctuations collapse and are larger than the one for the temperature fluctuations.

Clearly, there are multiple physical processes and thus scales at play near the top boundary at high $D$. This seems to be similar to the turbulent boundary layer with its inner and outer scales \citep{SMMARFM2011}. The behaviour of density can be analyzed to demarcate various regimes inside the boundary layers in compressible convection. We thus introduce the following boundary layer scales: (1) $\lambda_{T_{\rm max}}$ as the distance between the top wall and the location of the local maximum of the relative temperature fluctuations. (2) $\lambda_{\rho_{\rm max}}$ as the distance between the top wall and the maximum of the relative density fluctuations. Note that in OB and even anelastic convection,  $\lambda_{T_{\rm max}} = \lambda_{\rho_{\rm max}}$ \citep{jones2022}. Furthermore, (3) $\lambda_{\rho_{\rm min}}$ as the distance between the top wall and the location of the local minimum of the relative density fluctuations. This distance corresponds to the location where we start to observe significant compressibility effects. Finally, (4) $\lambda_{\rho_{{\rm max},T}}$ is the distance between the top wall and the local maximum of the relative density fluctuations, corresponding to the maximum of local temperature fluctuations. 

Next, we display all 4 scales $\lambda_X$ versus $D$ in the inset of figure \ref{fig:Fig3}(c). $\lambda_{T_{\rm max}}$ increases with $D$ which is consistent with the growth of the thermal boundary layer thickness at the top, $\lambda^{T}$, see figure \ref{fig:Fig2}(a). We see a sudden transition of $\lambda_{\rho_{\rm max}}$ at $D_{crit}= 0.65$ indicating that the maximum for $D>D_{\rm crit}$ is at $z=H$. Note also that for $D < 0.65$, $\lambda_{\rho_{\rm max}} \approx \lambda_{\rho_{{\rm max}, T}}$.  Both scales increase with $D$, but start to deviate from $\lambda_{T_{\rm max}}$ at $D < D_{\rm crit}= 0.65$. This deviation starts for $D \ge 0.5$; it suggests that pressure fluctuations are important already, but not as dominant as discussed before. It might thus indicate that the anelastic approximation is strictly speaking not applicable even for $D \ge 0.5$. Finally, $\lambda_{\rho_{\rm min}}$ starts from nearly zero at $D = 0.1$ with negligible compressibility. With increasing dissipation number $\lambda_{\rho_{\rm min}}$ increases.  This is a new top boundary layer regime induced by the increasing stratification. It is in line with an increase of $\lambda_{T_{\rm max}}$ and $\lambda^{T}$, see again the inset of figure \ref{fig:Fig2}(a).  We discuss the implications of this sublayer for the turbulent heat transfer in the next section. 

\section{Stratification impact on turbulent heat transfer}
\begin{figure}
%\begin{center}
\hspace{-10mm}
\includegraphics[clip,trim={0mm 0mm 0mm 0mm},width=1.2\textwidth]{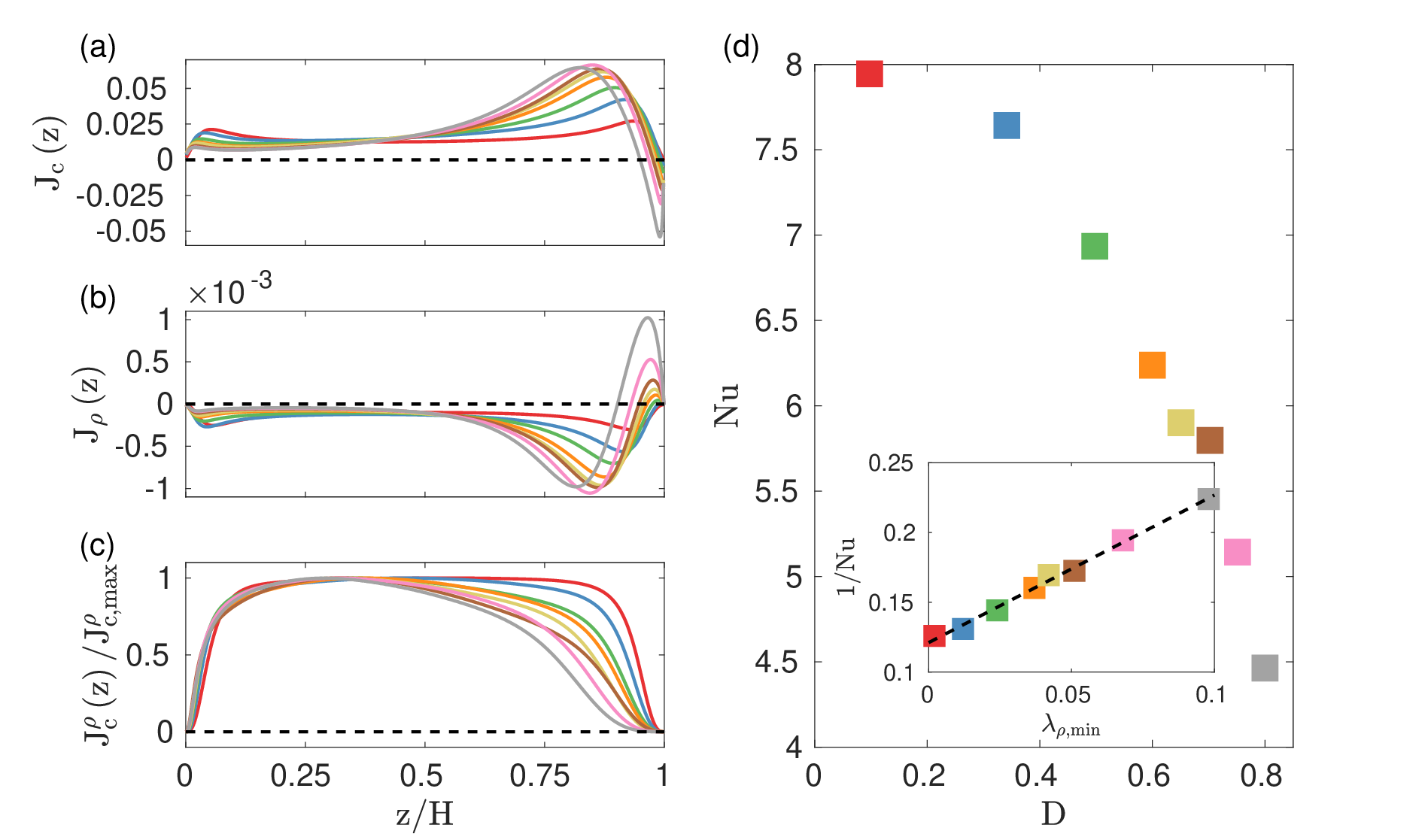}  
\caption{\label{fig:Fig4} 
Profiles of plane-time averaged stresses versus $D$. (a) Normalized convective  heat flux $J_{c}(z)=\langle u_z T^{\prime}(z)\rangle_{A,t}/[u_{z,{\rm rms}}(z)\langle T(z)\rangle_{A,t}]$. (b) Normalized density-temperature correlation $J_{\rho}(z)=\langle \rho^{\prime} T^{\prime}(z)\rangle_{A,t}/[\langle\rho(z)\rangle_{A,t}\langle T(z)\rangle_{A,t}]$. (c) Normalized density averaged convective heat flux where $J^{\rho}_{c}(z)=\langle \rho u_z T^{\prime}(z)\rangle_{A,t}$ and  (d) Nusselt number $Nu$ versus dissipation number $D$, see eq. \eqref{eq:Nu}. Inset of (c) displays $1/Nu$ versus $\lambda_{\rho_{min}}$. The dashed line corresponds to a fit $A + B \lambda_{\rho_{\rm min}}$ with coefficients $A=0.12$ and $B=1.05$. Line colouring corresponds to table \ref{tab:simu1}. Note that $T^{\prime}$ follows by a standard Reynolds decomposition, $T({\bm x},t)=\langle T(z)\rangle_{A,t}+T'({\bm x},t)$.} 
\begin{picture}(0,0)
\end{picture}
%\end{center}
\end{figure}

A systematic increase of the level of stratification, controlled by $D$, at relatively small superadibaticity of $\epsilon= 0.1$ has a significant impact on the normalized stress profiles as seen in figure \ref{fig:Fig4}. We show the variation of the normalized convective heat flux profiles $J_c(z)$ in (a) and the normalized density-temperature correlation profiles $J_{\rho}(z)$ in (b) for all 8 runs. They are qualitatively similar, but sign-flipped behaviour is observed for both second-order correlations in (a,b). An increasing sublayer with an average ``anti-convection"-behaviour, i.e., negative convective heat flux and positive density-temperature correlation can be detected. For $z\lesssim 0.9 \left(D= 0.8 \right)$ both mean profiles correspond to unstable convection. 

Next, we relate the depth of these stabilized regions to $\lambda_{\rho_{\rm min}}$ and thus to the regime transition at $D=D_{\rm crit}$. We therefore define three further distances or scales: (1) $\delta_{u_{z}T}$ as the width of negative normalized heat flux with respect to the top boundary. (2) $\lambda^{\rm max}_{\rho'T'}$ as the  distance from the top boundary to the location of the positive maximum of the normalized density-temperature correlation. (3) $\delta_{\rho'T'}$ as the width of the region of positive density-temperature correlations. 

%%%%%%%%Figure4Alert%%%%%
%%%%%%%%%%%%%%%%
These length scales along with $\lambda_{\rho_{\rm min}}$ are listed in table \ref{tab:simu1} for all DNS. We observe that for all runs, except for $D= 0.1$, the region of positive correlation exactly matches with the compressibility-dominated region, i.e., $\delta_{\rho'T'}=\lambda_{\rho_{\rm min}}$. Thus, there is also a direct proportionality between regions dominated by compressibility and anti-convection with $\delta_{{u_z}T} < \lambda_{\rho_{\rm min}}$. For $D \le 0.70$, we also find that $\delta_{{u_z}T} \approx \lambda^{\rm max}_{\rho'T'}$, the location of the maximum positive correlation. For $D > 0.7$, we get $\delta_{{u_z}T} > \lambda^{\rm max}_{\rho'T'}$, suggesting an increasing influence of compressibility. Furthermore, $\delta_{{u_z}T} < \lambda_{\rho_{\rm min}}$ implies that on average weak convective motions can occur at heights, $\lambda_{\rho_{\rm min}}\gtrsim z\gtrsim \delta_{{u_z}T}$, even though there are already significant compressibility effects. We recall for this discussion that even for the highest $D$, we still observe focused cold thermal plumes which detach from the top thus indicating locally positive convective flux. All these processes take places in the region where density fluctuations are tightly correlated to pressure fluctuations as we discussed in figure \ref{fig:Fig3}(c).  We also report that there is no trend of the location of maximum turbulent Mach number, $M^{\ast}_{t}=\max_z M_t(z)$, see figure \ref{fig:Fig2}(b), with $D$ even though $M_{t}^{\ast}$ increases with $D$. For all the cases considered, the distance of this maximum to the top wall is $H-z_{\ast} \in [0.16, 0.174]$, significantly larger than  $\lambda_{\rho_{\rm min}}$. This again suggests the inadequacy of $M_{t}$ alone to characterize general compressibility conditions \citep{DJ2020}. Particularly for the strongly asymmetric cases at highest $D$, the behaviour of compressible convection is completely different for same $M_{t}$-magnitude at either sides of $z_{\ast}$.  

\colb{In figure \ref{fig:Fig4}(c) we also plot the plane-averaged heat flux, $J^{\rho}_{c}(z)=\langle \rho' u_z T'\rangle_{A,t}$, normalized by the maximum value. We observe a region of practically zero convective heat flux near the top that increases with $D$. For $z>H-\delta_{u_{z}T}$, the flux $J^{\rho}_{c}$ is less than 1\% of its maximum.} Finally, we plot the Nusselt number defined as 
\begin{equation}
    Nu = \frac{1}{2} \left[\left . - \frac{H}{\epsilon  T_{B}} \frac{d \langle T_{sa} \rangle _{A,t}}{dz} \right|_{z=0}    \left . - \frac{H}{\epsilon  T_{B}} \frac{d \langle T_{sa} \rangle _{A,t}}{dz} \right|_{z=H}\right]
\label{eq:Nu}    
\end{equation}
with dissipation as function of $D$ in figure \ref{fig:Fig4}(d). The Nusselt number is calculated as the average of the heat flux at the top and bottom. Since the flows are statistically steady, the Nusselt number at the top and bottom are the same and the maximum mean Nusselt number  difference between the plate is found to be less than  $3 \%$  for all cases. Clearly visible is that the Nusselt number $Nu$ monotonically decreases with $D$. At critical $D= 0.65$, there seems to be slight  change in behaviour but is not significant. \colb{A decrease of $Nu$ with $D$ is again consistent with previous studies \citep{VerWSAJ2015,jones2022}. We plot the inverse of the Nusselt number versus $\lambda_{\rho_{\rm min}}$ in the inset of panel (d) and we observe that the trend can be reasonably approximated by a linear fit. Thus, we conclusively  demonstrate that an increasingly inefficient heat transfer is connected to structural changes in the top boundary layer.}

\section{Final discussion} 
The objective of the present study was to take the least complex, but fully compressible turbulent convection system to study the genuine effects of compressibility isolated from other major contributions to a non-Boussinesq behavior, such as temperature-dependent material properties, and to go beyond the Oberbeck-Boussinesq and anelastic limits. We took a configuration in the limit of strongly decreasing mean profiles of the thermodynamic state variables --- the high stratification regime. It corresponds to the dissipation number of $D\to 1$ for small superadiabaticity $\epsilon$. One prominent example for such a configuration might be the convection dynamics close to the solar surface. Even though the aim of our work is {\em not} to provide a realistic modeling of particularly this complex multi-physics flow, that would involve the coupling to magnetic fields and radiation transfer, the present simulations provide an insight into generic physical mechanisms which lead to highly asymmetric thermal plume motions, the fundamental local process in any convection flow that triggers fluid turbulence and transports heat. The limit cases of our simplified configuration lead already to a sparse network of thin plume structures which resembles similarities to the granule network at the Sun, see again figure \ref{fig:Fig1}(b). These coherent structures ``leak`` through the stratified top region and thus maintain a significantly reduced global turbulent heat transport. Future studies will aim at an increase of the aspect ratio and Rayleigh number together with a decrease of the Prandtl number and $T$-dependent material parameters.

\vspace{0.5cm}
\noindent
\textbf{Acknowledgements}

\noindent
J.P.J. is supported by grant no. SCHU 1410/31-1 of the Deutsche Forschungsgemeinschaft and a Postdoctoral Fellowship of the Alexander von Humboldt Foundation. The authors gratefully acknowledge the Gauss Centre for Supercomputing e.V. (https:// www.gauss-centre.eu) for funding this project by providing computing time through the John von Neumann Institute for Computing (NIC) on the GCS Supercomputer JUWELS at J\"ulich Supercomputing Centre (JSC). We thank Diego A. Donzis and Akansha Baranwal for sharing the cDNS code which has been adapted in this work \\

\noindent
The authors report no conflict of interest.

\bibliographystyle{jfm}
\bibliography{all}

\begin{thebibliography}{29}
\expandafter\ifx\csname natexlab\endcsname\relax\def\natexlab#1{#1}\fi
\def\au#1{#1} \def\ed#1{#1} \def\yr#1{#1}\def\at#1{#1}\def\jt#1{\textit{#1}}
  \def\bt#1{#1}\def\bvol#1{\textbf{#1}} \def\vol#1{#1} \def\pg#1{#1}
  \def\publ#1{#1}\def\arxiv#1{#1}\def\org#1{#1}\def\st#1{\textit{#1}}

\bibitem[Anders {\em et~al.\/}(2019)Anders, Lecoanet \& Brown]{AndersApJ2019}
{\sc \au{Anders, E.~H.}, \au{Lecoanet, D.} \& \au{Brown, B.~P.}} \yr{2019}
  \at{Entropy rain: Dilution and compression of thermals in stratified
  domains}.  \jt{Astrophys. J.}  \bvol{884}~(1),  \pg{65}.

\bibitem[Baranwal {\em et~al.\/}(2023{\natexlab{{\em a\/}}})Baranwal, Donzis \&
  Bowersox]{BDBaJFM2023}
{\sc \au{Baranwal, A.}, \au{Donzis, D.~A.} \& \au{Bowersox, R.~D.}}
  \yr{2023{\natexlab{{\em a\/}}}}  \at{Mach number and wall thermal boundary
  condition effects on near-wall compressible turbulence}.  \jt{arXiv preprint
  arXiv:2307.03265} .

\bibitem[Baranwal {\em et~al.\/}(2023{\natexlab{{\em b\/}}})Baranwal, Donzis \&
  Bowersox]{BDBaiaa2023}
{\sc \au{Baranwal, A.}, \au{Donzis, D.~A.} \& \au{Bowersox, R.~D.}}
  \yr{2023{\natexlab{{\em b\/}}}} Turbulent heat flux in supersonic flows for
  different thermal boundary conditions.  \bt{In {\em AIAA SCITECH 2023
  Forum\/}},  \pg{p. 0868}.

\bibitem[Baranwal {\em et~al.\/}(2022)Baranwal, Donzis \& Bowersox]{BDBjfm2022}
{\sc \au{Baranwal, A.}, \au{Donzis, D.~A.} \& \au{Bowersox, R. D.~W.}}
  \yr{2022}  \at{Asymptotic behaviour at the wall in compressible turbulent
  channels}.  \jt{J. Fluid. Mech.}  \bvol{933}.

\bibitem[Brandenburg(2016)]{BrandenburgApJ2017}
{\sc \au{Brandenburg, A.}} \yr{2016}  \at{Stellar mixing length theory with
  entropy rain}.  \jt{Astrophys. J.}  \bvol{832}~(1),  \pg{6}.

\bibitem[Chill{\`a} \& Schumacher(2012)]{CJ2012}
{\sc \au{Chill{\`a}, F.} \& \au{Schumacher, J.}} \yr{2012}  \at{{New
  perspectives in turbulent Rayleigh-B{\'e}nard convection}}.  \jt{Eur. Phys.
  J. E}  \bvol{35}~(7),  \pg{1--25}.

\bibitem[Christensen-Dalsgaard(2002)]{christ2002}
{\sc \au{Christensen-Dalsgaard, J.}} \yr{2002}  \at{Helioseismology}.  \jt{Rev.
  Mod. Phys.}  \bvol{74}~(4),  \pg{1073}.

\bibitem[Cossette \& Rast(2016)]{CossetteRast2016}
{\sc \au{Cossette, J.-F.} \& \au{Rast, M.~P.}} \yr{2016}  \at{{Supergranulation
  as the largest buoyantly driven convective scale of the Sun}}.
  \jt{Astrophys. J. Lett.}  \bvol{829}~(1),  \pg{L17}.

\bibitem[Donzis \& John(2020)]{DJ2020}
{\sc \au{Donzis, D.~A.} \& \au{John, J.~P.}} \yr{2020}  \at{Universality and
  scaling in homogeneous compressible turbulence}.  \jt{Phys. Rev. Fluids}
  \bvol{5},  \pg{084609}.

\bibitem[Hanasoge {\em et~al.\/}(2012)Hanasoge, Duvall~Jr \&
  Sreenivasan]{HanasogePNAS2012}
{\sc \au{Hanasoge, S.~M.}, \au{Duvall~Jr, T.~L.} \& \au{Sreenivasan, K.~R}}
  \yr{2012}  \at{Anomalously weak solar convection}.  \jt{Proc. Nat. Acad. Sci.
  USA}  \bvol{109}~(30),  \pg{11928--11932}.

\bibitem[Hanson {\em et~al.\/}(2020)Hanson, Duvall, Birch, Gizon \&
  Sreenivasan]{HansonAA2020}
{\sc \au{Hanson, C.~S.}, \au{Duvall, T.~L.}, \au{Birch, A.~C.}, \au{Gizon, L.}
  \& \au{Sreenivasan, K.~R}} \yr{2020}  \at{Solar east-west flow correlations
  that persist for months at low latitudes are dominated by active region
  inflows}.  \jt{Astron. Astrophys.}  \bvol{644},  \pg{A103}.

\bibitem[Jeffreys(1930)]{jeff1930}
{\sc \au{Jeffreys, H.}} \yr{1930} The instability of a compressible fluid
  heated below.  \bt{In {\em Math. Proc. Camb. Phil. Soc.\/}}, ,
  \vol{vol.~26},  \pg{pp. 170--172}. Cambridge University Press.

\bibitem[John {\em et~al.\/}(2019)John, Donzis \& Sreenivasan]{JDS2019}
{\sc \au{John, J.~P.}, \au{Donzis, D.~A} \& \au{Sreenivasan, K.~R}} \yr{2019}
  \at{Solenoidal scaling laws for compressible mixing}.  \jt{Phys. Rev. Lett.}
  \bvol{123}~(22),  \pg{224501}.

\bibitem[John {\em et~al.\/}(2020)John, Donzis \& Sreenivasan]{JDScst2020}
{\sc \au{John, J.~P.}, \au{Donzis, D.~A.} \& \au{Sreenivasan, K.~R}} \yr{2020}
  \at{Compressibility effects on the scalar dissipation rate}.  \jt{Combust.
  Sci. Technol.}  \bvol{192}~(7),  \pg{1320--1333}.

\bibitem[John {\em et~al.\/}(2021)John, Donzis \& Sreenivasan]{JDSJFM2021}
{\sc \au{John, J.~P.}, \au{Donzis, D.~A.} \& \au{Sreenivasan, K.~R.}} \yr{2021}
   \at{Does dissipative anomaly hold for compressible turbulence?}  \jt{J.
  Fluid. Mech.}  \bvol{920},  \pg{{A20}}.

\bibitem[John \& Schumacher(2023)]{JPSc2023}
{\sc \au{John, J.~P.} \& \au{Schumacher, J.}} \yr{2023}  \at{Strongly
  superadiabatic and stratified limits of compressible convection}.  \jt{arXiv
  preprint arXiv:2302.03621} .

\bibitem[Jones {\em et~al.\/}(2022)Jones, Mizerski \& Kessar]{jones2022}
{\sc \au{Jones, C.~A.}, \au{Mizerski, K.~A} \& \au{Kessar, M.}} \yr{2022}
  \at{Fully developed anelastic convection with no-slip boundaries}.  \jt{J.
  Fluid Mech.}  \bvol{930},  \pg{A13}.

\bibitem[Lele(1992)]{lele1992}
{\sc \au{Lele, S.~K.}} \yr{1992}  \at{Compact finite-difference schemes with
  spectral-like resolution}.  \jt{J. Comp. Phys.}  \bvol{103},  \pg{16--42}.

\bibitem[Magic {\em et~al.\/}(2013)Magic, Collet, Asplund, Trampedach, Hayek,
  Chiavassa, Stein \& Nordlund]{magic2013a}
{\sc \au{Magic, Z.}, \au{Collet, R.}, \au{Asplund, M.}, \au{Trampedach, R.},
  \au{Hayek, W.}, \au{Chiavassa, A.}, \au{Stein, R.~F.} \& \au{Nordlund,
  {\AA}.}} \yr{2013}  \at{{The Stagger-grid: A grid of 3D stellar atmosphere
  models--I. Methods and general properties}}.  \jt{Astron. Astrophys.}
  \bvol{557},  \pg{A26}.

\bibitem[Nordlund {\em et~al.\/}(2009)Nordlund, Stein \&
  Asplund]{nordlund2009solar}
{\sc \au{Nordlund, {\AA}.}, \au{Stein, R.~F.} \& \au{Asplund, M.}} \yr{2009}
  \at{Solar surface convection}.  \jt{Living Rev. Solar Phys.}  \bvol{6}~(1),
  \pg{1--117}.

\bibitem[Ogura \& Phillips(1962)]{OP1962}
{\sc \au{Ogura, Y.} \& \au{Phillips, N.~A.}} \yr{1962}  \at{Scale analysis of
  deep and shallow convection in the atmosphere}.  \jt{J. Atmos. Sci}
  \bvol{19}~(2),  \pg{173--179}.

\bibitem[Rast(1998)]{rast1998}
{\sc \au{Rast, M.~P.}} \yr{1998}  \at{Compressible plume dynamics and
  stability}.  \jt{J. Fluid Mech.}  \bvol{369},  \pg{125--149}.

\bibitem[Rincon \& Rieutord(2018)]{Rincon2018}
{\sc \au{Rincon, F.} \& \au{Rieutord, M.}} \yr{2018}  \at{{The Sun’s
  supergranulation}}.  \jt{Living Rev. Solar Phys.}  \bvol{15},  \pg{1--74}.

\bibitem[Scheel \& Schumacher(2014)]{JSJSJFM2014}
{\sc \au{Scheel, J.~D.} \& \au{Schumacher, J.}} \yr{2014}  \at{{Local boundary
  layer scales in turbulent Rayleigh--B{\'e}nard convection}}.  \jt{J. Fluid.
  Mech.}  \bvol{758},  \pg{344--373}.

\bibitem[Schumacher \& Sreenivasan(2020)]{SKRMPhys2020}
{\sc \au{Schumacher, J.} \& \au{Sreenivasan, K.~R.}} \yr{2020}
  \at{{Colloquium: Unusual dynamics of convection in the Sun}}.  \jt{Rev. Mod.
  Phys.}  \bvol{92}~(4),  \pg{041001}.

\bibitem[Smits {\em et~al.\/}(2011)Smits, McKeon \& Marusic]{SMMARFM2011}
{\sc \au{Smits, A.~J.}, \au{McKeon, B.~J.} \& \au{Marusic, I.}} \yr{2011}
  \at{{High--Reynolds number wall turbulence}}.  \jt{Annu. Rev. Fluid Mech.}
  \bvol{43},  \pg{353--375}.

\bibitem[Tilgner(2011)]{tilgner2011}
{\sc \au{Tilgner, A.}} \yr{2011}  \at{{Convection in an ideal gas at high
  Rayleigh numbers}}.  \jt{Phys. Rev. E.}  \bvol{84}~(2),  \pg{026323}.

\bibitem[Verhoeven {\em et~al.\/}(2015)Verhoeven, Wieseh{\"o}fer \&
  Stellmach]{VerWSAJ2015}
{\sc \au{Verhoeven, J.}, \au{Wieseh{\"o}fer, T.} \& \au{Stellmach, S.}}
  \yr{2015}  \at{{Anelastic versus fully compressible turbulent
  Rayleigh--B{\'e}nard convection}}.  \jt{Astrophys.J.}  \bvol{805}~(1),
  \pg{62}.

\bibitem[Wu \& Libchaber(1991)]{WL1991}
{\sc \au{Wu, X.-Z.} \& \au{Libchaber, A.}} \yr{1991}  \at{{Non-Boussinesq
  effects in free thermal convection}}.  \jt{Phys. Rev. A}  \bvol{43}~(6),
  \pg{2833--2839}.

\end{thebibliography}

\end{document}